\documentclass[11pt]{article}
\usepackage[margin=1in]{geometry}
\usepackage{setspace}
\usepackage[utf8]{inputenc}
\usepackage{microtype}
\usepackage{amsmath,amssymb,amsthm,mathtools}
\usepackage{booktabs,needspace,enumitem,array,float}
\usepackage[hidelinks]{hyperref}
\usepackage{cleveref}

\newtheorem{theorem}{Theorem}[section]
\newtheorem{lemma}[theorem]{Lemma}
\newtheorem{proposition}[theorem]{Proposition}
\newtheorem{corollary}[theorem]{Corollary}

\theoremstyle{definition}

\newcommand{\F}{\mathbb{F}}
\newcommand{\GL}{\operatorname{GL}}
\newcommand{\rank}{\operatorname{rank}}
\newcommand{\E}{\mathbb{E}}
\newcommand{\Var}{\operatorname{Var}}
\newcommand{\eps}{\varepsilon}
\newcommand{\Inv}{\mathcal{I}}
\newcommand{\Det}{\mathcal{D}}
\newcommand{\Mult}{\mathcal{M}}

\title{Worst-Case to Average-Case Reductions for Basic Linear Algebra}
\author{Or Zamir\\\emph{Tel Aviv University}}
\date{}
\begin{document}
\maketitle
\begin{abstract}
We study worst-case to average-case reductions for basic linear-algebraic problems over finite fields. 
A seminal work of Blum, Luby, and Rubinfeld established that any algorithm that successfully multiplies more than~$75\%$ of matrix pairs over any finite field, implies an algorithm with essentially the same running time multiplying any pair of matrices. A sequence of follow-ups reduced the needed fraction of successful input pairs to any positive constant.
Despite most of the other basic linear algebraic operations being computationally equivalent to matrix multiplication in the worst case, no such worst-case to average-case reductions were known for them.
Recently, Vaikuntanathan and Zamir introduced a cryptographic primitive called trapdoor matrices and used them to prove such worst-case to average-case reductions for matrix inversion, solving a linear system, and computing the determinant, all conditioned on common cryptographic assumptions. 

Our main result is the first \emph{unconditional} time-preserving worst-case to average-case reduction for matrix inversion. Any algorithm that correctly inverts any constant positive fraction of uniformly random invertible matrices implies a worst-case such algorithm.
We also significantly strengthen the conditional reductions based on trapdoored matrices: We establish reductions for deciding invertibility and computing rank at the optimal success thresholds, and qualitatively improve the success threshold needed in the reduction for determinant computation---while previous reductions all required the success probability to be at least~$1/2$, we show that \emph{any} fixed positive average-case success probability suffices for exact worst-case determinant computation over infinitely many finite fields.
\end{abstract}

\section{Introduction}\label{sec:introduction}

Matrix multiplication, matrix inversion, solving linear systems, rank, and determinants are among the foundational operations of computation. Their worst-case running times are closely connected: fast matrix multiplication yields fast algorithms for the other problems through classical reductions and matrix decompositions~\cite{Strassen1969,BunchHopcroft1974,IbarraMoranHui1982,KellerGehrig1985}. 
The analogous reverse equivalence is also known for many of these problems in various computational models~\cite{BaurStrassen1983,DoneyMusco2025}.

Worst-case performance, however, is only part of the picture, especially today when large-scale linear algebra is central to real-life computational workloads and machine learning. 
Average-case algorithms can potentially be much faster, and the usual worst-case reductions pass through highly structured matrices on which a uniform-input algorithm need not be reliable. 
It is therefore natural to ask whether the classical relationships persist in the form of \emph{worst-case to average-case reductions} that preserve the running-time exponent. 

The seminal work of Blum, Luby, and Rubinfeld~\cite{BlumLubyRubinfeld1993} established this phenomenon for matrix multiplication over finite fields, initially under a high average-case success assumption. Subsequent work progressively improved the tolerated error, culminating in reductions from arbitrarily small positive success or agreement under appropriate formulations~\cite{AsadiGolovnevGurShinkar2022,GolaShinkarSingh2024,HiraharaShimizu2025STOC,HiraharaShimizu2025ICALP,ShinkarSingh2025}. Despite the worst-case relationships among the underlying problems, comparable uniform-input reductions were not known for full inversion, solving a linear system, rank, or determinant. 
Earlier efforts to resolve this gap stopped short of resulting in a true worst-case to average-case reduction: Program-self-correction results for these tasks either used auxiliary computational routines or imposed correctness on additional, structured input distributions~\cite{BlumLubyRubinfeld1993,GoldwasserEtAl2008,Rothblum2009}; these guarantees differ significantly from correctness solely on uniformly random matrices.

Vaikuntanathan and Zamir~\cite{VaikuntanathanZamir2026} recently introduced \emph{trapdoored matrices}: computationally pseudorandom matrices that admit near-linear-time matrix-vector multiplication given a secret trapdoor. 
As an application, they obtained the first (albeit conditional) worst-case to average-case reductions for matrix inversion, solving a linear system, and determinant, based on standard cryptographic assumptions such as Learning Parity with Noise. 
This raised two questions: can the assumptions be removed, and how little success on the uniform distribution suffices? 
For determinant, Hhan and Lee~\cite{HhanLee2026} then obtained better success thresholds by constructing special trapdoored matrices whose determinants are known, using a stronger, more specialized cryptographic assumption.

We answer the first question \emph{unconditionally} for matrix inversion, and substantially improve the conditional picture for rank, invertibility, and determinant. 


\subsection{Unconditional worst-case to average-case reduction for matrix inversion}
Our first contribution is the first unconditional such reduction for inversion. 
\begin{theorem}[Worst-case multiplication from average-case inversion]\label{thm:main-inverse}
Let $\F_q$ be a finite field and $\eps>0$. 
Suppose that for all sufficiently large $m$, a randomized algorithm $\Inv_m$, running in some non-decreasing time $T(m)\geq m^2$, satisfies
\[
 \Pr_{Y\sim\GL_m(\F_q)}[\Inv_m(Y)=Y^{-1}]\ge\eps.
\]
Then any matrices $B,C\in\F_q^{n\times n}$ can be multiplied, with high probability, in time
\[
 \widetilde O\bigl(\;\frac{1}{\varepsilon}\;T(2n)\;\bigr).
\]
\end{theorem}

Thus an inverter succeeding on any constant fraction of random invertible matrices already captures the exponent of worst-case matrix multiplication, uniformly over finite fields. 
The proof combines two steps: (1) a new identity involving \emph{complementary Schur complements}, which suffices for a reduction when the success probability is a constant close to one, and (2) an error-amplification procedure based on low-rank perturbations and the spectral gap of a rank-one walk, which amplifies any positive constant success probability to the one necessary for the first step. 

The resulting worst-case matrix multiplication algorithm yields worst-case full inversion and the other standard tasks via classical algorithms~\cite{Strassen1969,BunchHopcroft1974,IbarraMoranHui1982,KellerGehrig1985}. 
The dependence on $\eps$ in this reduction is optimal.

\subsection{Conditional reductions for rank and invertibility at the optimal thresholds}
Our remaining results assume the existence of the pseudorandom trapdoored-matrix primitive of~\cite{VaikuntanathanZamir2026, braverman2025practical}, which they proved follows from standard cryptographic assumptions such as Learning Parity with Noise or McEliece. 

We first observe that generically, any trapdoor matrix family gives more than fast matrix-vector products: Wiedemann's method, together with randomized preconditioning and Berlekamp--Massey, allows computing the determinant or solve an individual invertible system of the trapdoor matrix in near-quadratic time~\cite{Wiedemann1986,ChenEberlyKaltofenSaundersTurnerVillard2002}. 
In particular, we immediately recover the improved determinant thresholds of Hhan and Lee~\cite{HhanLee2026} using the generic trapdoor assumption of~\cite{VaikuntanathanZamir2026}.

We then use these primitives to construct reductions at the natural baselines for predicting invertibility and rank. 
Denote by
\[
 s_{q,n}=\Pr_{X\sim\F_q^{n\times n}}[\det X=0]
 =1-\prod_{i=1}^{n}(1-q^{-i})\leq \frac{1}{q-1}
\]
the singularity probability of a uniform matrix. Let $s=s_{q,n}$. 
For invertibility, the trivial algorithm that simply returns a fixed constant answer succeeds with probability $\max\{s,1-s\}$; for exact rank, a corresponding baseline is the probability of the most common rank.
\begin{theorem}[Conditional rank and invertibility reductions]\label{thm:rank-conditional}
Fix a finite field $\F_q$ and $\eps>0$, and assume there is a trapdoored-matrix family with near-linear-time vector action and near-quadratic-time sampling. 
Suppose a polynomial-time algorithm $\mathcal A$, running in time $T(n)$, either
\begin{enumerate}[label=(\roman*),topsep=3pt,itemsep=2pt]
\item decides invertibility of a uniform $n\times n$ matrix with probability $\geq \max\{s_{q,n},1-s_{q,n}\}+\eps$, or
\item computes its rank with probability at least $\max_r\Pr[\rank X=r]+\eps$.
\end{enumerate}
Then, worst-case invertibility can be decided with high probability in $\widetilde O_q(\eps^{-2}(T(n)+n^2))$ time, and exact worst-case rank follows in $\widetilde O_q(\eps^{-2}(T(2n)+n^2))$ time. 
\end{theorem}
The thresholds are optimal as they require being only slightly better than the best constant predictors.

\subsection{An hierarchy of conditional reductions for determinant}
Determinant admits a richer hierarchy. 
Our starting point is the simple reduction presented by~\cite{VaikuntanathanZamir2026} using a trapdoor matrix family:
Draw a trapdoored matrix~$R$, then use the average-case determinant algorithm twice to compute the quotient~$\Det(RA)/\Det(R)$, where the product~$RA$ can be quickly computed using the trapdoor.
As both arguments given to~$\Det$ are pseudo-random, if its success probability is high enough then we would frequently see the correct value of~$\det(A)$ appearing as the quotient.
This requires the success probability to be larger than~$\frac{3}{4}$, as we would want that at least~$\frac{1}{2}$ of such attempts would return the correct determinant as it would then appear as the majority output of all attempts. 
That success probability threshold was improved to roughly~$\frac{1}{2}$ in~\cite{HhanLee2026} by designing a trapdoor matrix family for which computing the determinant is easy, and thus the second oracle call~$\Det(R)$ becomes unnecessary. 
Our aforementioned observation that in \emph{every} trapdoor matrix family determinant computation is quick is sufficient to recover this result from less specialized cryptographic assumptions.

Unfortunately, the threshold being above~$\frac{1}{2}$ appears as an inherent barrier: 
Let~$\F_q$ be a field with characteristic strictly larger than two, and consider the algorithm~$\Det(A)$ that correctly computes~$\det(A)$ but then returns it multiplied with a random sign from~$\{-1,1\}$.
Then,~$\Det$ succeeds with probability~$s_{q,n}+\frac{1-s_{q,n}}{2}>\frac{1}{2}$.
On the other hand, we are currently unaware of any efficient reductions that preserve a non-multiplicative function of the determinant; in particular, we cannot disambiguate~$\det(A)$ from~$-\det(A)$ efficiently.
The same holds for any ambiguity up to a subgroup~$H\leq \F_q^\times$.
The same barrier also came up in the context of worst-case reductions within linear algebra: while many of the basic operations are known to be equivalent, it is still unknown whether computing rank or invertibility are as hard as computing the determinant or matrix multiplication (see e.g.~\cite{DoneyMusco2025}). An invertibility oracle can be viewed as an algorithm that recovers the determinant up to the subgroup~$H=\F_q^\times$, yet we do not know how to use such an oracle to recover the determinant.

Our main result on determinant computation is that the aforementioned barrier is essentially the only roadblock: we show that if the average-case algorithm is even slightly better than the constant algorithm always returning zero, it already can be used to solve invertibility in the worst-case.
Furthermore, if the algorithm is slightly better than what an invertibility algorithm implies (that is, return either~$0$ or~$1$ according to whether~$\det(A)=0$), then it already suffices to recover~$\det(A)$ up to multiplication with some small subgroup~$H<\F_q^\times$.
In particular, if we consider finite fields~$\F_{2^k}$ of characteristic two, then the size of the smallest non-trivial such subgroup is often far larger than two; resulting in thresholds that are vanishing.

The above establishes two interesting thresholds:
The always-zero predictor succeeds with probability $s_{q,n}$. 
An algorithm that instead of computing the determinant merely solves invertibility, and then returns zero on singular matrices and one on invertible matrices has success probability $s_{q,n}+(1-s_{q,n})/(q-1)$. 
\begin{theorem}[Conditional determinant recovery]\label{thm:det-hierarchy}
Under the same trapdoored-matrix assumption, let $\Det$ be a polynomial-time determinant algorithm running in time $T(n)$.
\begin{enumerate}[label=(\roman*),topsep=3pt,itemsep=2pt]
\item If $\Det$ is correct on at least an $s_{q,n}+\eps$ fraction of uniform matrices, then worst-case invertibility is decidable in $\widetilde O_q(\eps^{-2}(T(n)+n^2))$ time, and exact rank follows in $\widetilde O_q(\eps^{-2}(T(2n)+n^2))$ time.
\item Suppose $q>2$ and, on a uniform \emph{invertible} matrix $Y$,
\[
 p:=\Pr[\Det(Y)=\det Y]\ge\frac1{q-1}+\eps.
\]
There is a randomized preprocessing procedure, depending only on~$\Det$, which produces a proper subgroup $H<\F_q^\times$. Subsequently, for every $A$, the reduction computes the coset $\det(A)H$ with high probability, in $\widetilde O_{q,\eps}(T(n)+n^2)$ time. The subgroup has order $h\mid(q-1)$ with
\[
 h<\frac1{p-\eps/4}<\min\{q-1,2/\eps\}.
\]
Equivalently, the reduction exactly computes $(\det A)^h$.
\end{enumerate}
\end{theorem}
The subgroup $H$ can depend on the oracle and on the randomness used during calibration, but is fixed before seeing any input matrix~$A$. 
The second hypothesis follows, in particular, from overall determinant accuracy $s_{q,n}+(1-s_{q,n})/(q-1)+\eps$; but we only use the weaker conditional-on-invertibility hypothesis.

We note that the size of the subgroup~$H$ is bounded by a constant~$O(1/\varepsilon)$ for any finite field, and thus we always obtain a recovery of the determinant up to a constant-sized list of options.
The same subgroup result also improves \emph{exact} determinant recovery for every even field of order at least four. 

\begin{corollary}[Exact determinant recovery]\label{thm:det-reduction}
Denote by $\ell$ the smallest prime divisor of $q-1$.
Under the generic trapdoor assumption, exact worst-case determinant computation follows at the same running-time exponent from average success at least
\[
\begin{cases}
 s_{2,n}+\eps, &q=2,\\[1mm]
 s_{q,n}+\dfrac{1-s_{q,n}}{\ell}+\eps, &q>2.
\end{cases}
\]
More generally, for $q>2$ it suffices to succeed on a fraction $1/\ell+\eps$ of uniform invertible matrices. 
\end{corollary}
Note that for odd $q$, $\ell=2$, recovering the previous $(1+s_{q,n})/2+\eps$ threshold of~\cite{HhanLee2026} but under the general trapdoor assumption. 
For every even $q\ge4$, we have $\ell\ge3$, giving a strictly lower threshold.
Furthermore, we observe that we can choose fields for which the \emph{smallest} prime divisor of~$q-1$ grows to infinity.
\begin{corollary}[Arbitrarily small average-case success over infinitely many fields]\label{cor:small-success}
For every fixed $\varepsilon>0$, there are infinitely many finite fields $\F_q$ for which, assuming the corresponding trapdoored-matrix family, any polynomial-time algorithm that computes the determinant correctly on at least an $\varepsilon$ fraction of uniform matrices yields an exponent-preserving algorithm for the worst-case determinant. 
\end{corollary}
The above is obtained, for example, by picking $q=2^r$ for all sufficiently large odd prime exponents $r$. For these fields the success threshold is $O(1/\log q)$.
Previously known sufficient thresholds for exact determinant all exceeded $1/2$ over every finite field.

\begin{table}[b]
\centering\small
\renewcommand{\arraystretch}{1.09}
\begin{tabular}{@{}>{\raggedright\arraybackslash}p{0.47\linewidth}>{\raggedright\arraybackslash}p{0.41\linewidth}@{}}
\toprule
Average-case hypothesis & Worst-case consequence \\
\midrule
Full inversion: success $\eps$ on uniform $\GL_n$ & Matrix multiplication; \emph{unconditional} \\
Invertibility: success above most-likely response & Invertibility and rank \\
Rank: success above most-likely response & Invertibility and rank \\
Determinant: success above probability of~$0$ & Invertibility and rank \\
Determinant: invertible-input success $1/(q-1)+\eps$ & Determinant modulo a small subgroup \\
\bottomrule
\end{tabular}
\caption{Main reductions. All rows except full inversion assume the generic trapdoored-matrix hypothesis.}\label{tab:results}
\end{table}

\subsection{Organization}
Section~\ref{sec:unconditional-inversion} establishes the unconditional inversion theorem. Section~\ref{sec:rank} proves the generic trapdoor capabilities and the conditional rank and invertibility reductions. Section~\ref{sec:det} develops the determinant results. 
Finally, we follow with limitations and related open problems in Section~\ref{sec:discussion}. 

\paragraph{Acknowledgments and AI Usage:}
The author deeply thanks Mark Braverman, Dan Karliner, and Vinod Vaikuntanathan for helpful discussions on these problems. 
Most core ideas as well as the framing and final versions of all reductions and corollaries are biologically-generated by the author. 
Publicly-available AI tools were extensively used as a significant speed multiplier: to quickly test directions, to fill technical gaps, to improve dependence on parameters, to verify each introduced primitive is used to exhaustion, etc. 

\newpage
\section{Unconditional reductions for matrix inversion}\label{sec:unconditional-inversion}
Our reduction consists of two distinct parts.

In Section~\ref{subsec:inverse-high} we establish a reduction that works when the average-case algorithm has a very large (albeit constant) success probability, say,~$99.999999\%$.
In hindsight, the reduction is rather simple, and leverages combining insights from both worst-case reductions within linear algebraic tasks and the previous worst-case to average-case reduction for matrix multiplication.
In the former, we usually embed a linear algebraic task inside the result of a different operation on a larger block-matrix. For example, we can show that inversion implies general matrix multiplication via the following identity:
\[
\begin{pmatrix}I&A&0\\ 0&I&B\\0&0&I\end{pmatrix}^{-1}
=
\begin{pmatrix}I&-A&AB\\ 0&I&-B\\0&0&I\end{pmatrix}
.
\]
The problem with using such reductions with an average-case algorithm is that the queried matrix is very structured, and thus does not look like a uniform matrix. 
To overcome this, we design a more complicated identity that requires several sequential inversions to finally recover the product~$AB$.
We show that if~$A$ and~$B$ themselves are uniform, then all inversion calls made in that reduction are also uniform. This implies that average-case inversion implies average-case matrix multiplication, which we already know via previous reductions that also implies worst-case matrix multiplication (and hence also inversion).

In Section~\ref{sec:inversion-amplification} we amplify any constant success probability~$\varepsilon>0$ to our desired high constant probability with an overhead that is near-linear in~$1/\varepsilon$.
This part is more technical than the first step, yet conceptually it is rather intuitive:
Consider a rank-one perturbation~$A\rightarrow A+u^tv$ for random vectors~$u,v$.
A classical linear algebraic identity by Sherman--Morrison (which is a special case of the Woodbury formula) shows that the inverses of the matrix before and after the perturbation can be quickly computed from each other:
\[
(A+u^tv)^{-1}=A^{-1} - \frac{(A^{-1}u)(v^t A^{-1})}{1+v^t A^{-1} u}.
\]
Using relatively basic Fourier analysis, we show that this perturbation ``gets us closer'' to any constant-density event on the set of all matrices. In particular, a random walk consisting of sufficiently many such steps gets us to a matrix that has roughly the same success probability in the average-case algorithm as a truly uniform matrix.

\subsection{Reduction from high-success average inversion}\label{subsec:inverse-high}
We first prove Theorem~\ref{thm:main-inverse} assuming high average success. For this proof, write $\gamma_j=1-s_{q,j}=\prod_{i=1}^{j}(1-q^{-i})\ge1/4$. All estimates in this section are uniform in the field size $q$. We assume that $n\ge3$.

\begin{proposition}\label{prop:high-success}
Suppose that for $m=n,2n$ the algorithm $\Inv_m$ runs in time $T(m)$ and correctly inverts a uniformly random matrix on $\GL_m(\F_q)$ with probability at least $1-10^{-6}$. Then any two matrices in $\F_q^{n\times n}$ can be multiplied with high probability in time $\widetilde O(T(n)+T(2n)+n^2)$.
\end{proposition}

\paragraph{Complementary Schur complements.}
Let $B,C\in\F_q^{n\times n}$ be the two matrices whose product we want to compute. Sample independent uniform $R,A,A'\in\F_q^{n\times n}$ and use the inversion algorithm to try computing $D=R^{-1}$ and $D'=(I-R)^{-1}$. We next form the block matrices
\[
M=\begin{pmatrix}A&B\\ C&D\end{pmatrix},\qquad
M'=\begin{pmatrix}A'&B\\ C&D'\end{pmatrix},
\]
call the inverter on both, and denote by $H,H'$ the upper-left $n\times n$ blocks of its answers. Finally, call the inverter on $H$ and $H'$ and return
\[
Q=A+A'-H^{-1}-(H')^{-1}.
\]
Thus the procedure makes six inversion calls, four on matrices of dimension $n$ and two on matrices of dimension $2n$. If an intermediate answer is incorrect (either because of an error introduced by the inverter or by the input matrix simply not being invertible), the procedure returns an arbitrary matrix.

Recall the Schur complement formula (see, e.g., \cite{Zhang2005}).
For a block matrix
\[
M=\begin{pmatrix} A&B\\ C&D\end{pmatrix},
\]
where $D$ is invertible, the Schur complement of $D$ is
$S=A-BD^{-1}C$. If $S$ is also invertible, then
\[
M^{-1}=
\begin{pmatrix}
S^{-1} & -S^{-1}BD^{-1}\\
-D^{-1}CS^{-1} & D^{-1}+D^{-1}CS^{-1}BD^{-1}
\end{pmatrix}.
\]
In particular, the upper-left block of $M^{-1}$ is
$(A-BD^{-1}C)^{-1}$.

To analyze the ideal execution, write $S=A-BRC$ and $S'=A'-B(I-R)C$, and let $\mathcal E$ be the event that $R,I-R,S,S'$ are all invertible. On $\mathcal E$, the Schur complement formula gives $H=S^{-1}$ and $H'=(S')^{-1}$. Therefore, whenever the six oracle calls are correct,
\[
Q=A+A'-S-S'=BRC+B(I-R)C=BC.
\]
Note that beyond the six inversion calls, we only add matrices or their sub-matrices, and never perform any multiplication ourselves.

Let $\beta_n=\Pr_R[R,I-R\in\GL_n(\F_q)]$. We will prove in Lemma~\ref{lem:complementary} that $\beta_n\ge1/32$ for $n\ge2$ over every finite field. For every fixed $B,C$ and every $R$ with $R,I-R$ invertible, the independently uniform matrices $A,A'$ make $S,S'$ independent uniform matrices. It follows that
\[
\Pr[\mathcal E\mid B,C]=\beta_n\gamma_n^2\ge\frac1{512}.
\]

We next bound the probability that an ideal inversion call is incorrect. This probability will be averaged over uniform independent $B,C$: for a fixed pair $B,C$, the matrices supplied to later calls need not be uniformly distributed, yet as we already know from previous reductions, it suffices to solve matrix multiplication in the average-case.

\begin{lemma}\label{lem:ideal-inputs}
Suppose the inverter fails on at most a $10^{-6}$ fraction
of uniformly random invertible matrices, in dimensions
$n$ and $2n$. Then the probability that $\mathcal E$
holds but one of the six inversion calls in our
construction is incorrect is at most $6\cdot 10^{-6}$.
\end{lemma}

\begin{proof}
We analyze the six calls assuming that all preceding
calls returned the correct inverses. If all six calls
are correct under this assumption, then the actual
execution follows the same sequence and returns $BC$.

For the first two calls, on $R$ and $I-R$, the claim
is immediate: both matrices are uniformly random before
conditioning on $\mathcal E$, and on $\mathcal E$ they
are invertible. Thus the probability that $\mathcal E$
holds and either particular call fails is at most
$10^{-6}$.

For the calls on $M$ and $M'$, observe that each is
distributed as a uniformly random $2n\times 2n$ matrix,
except that samples with a singular bottom-right block
are discarded (without resampling). Indeed, inversion
permutes the invertible matrices, and $I-R$ is uniform
whenever $R$ is uniform. Since $\mathcal E$ guarantees
that $M$ and $M'$ are invertible, discarding samples
cannot increase the probability that either call fails
together with $\mathcal E$. Thus each call contributes
at most $10^{-6}$ to the failure probability.

Finally, consider the calls on $H=S^{-1}$ and
$H'=(S')^{-1}$. For any fixed $B,C,R$, the
independent uniform choices of $A,A'$ make
\[
S=A-BRC,\qquad S'=A'-B(I-R)C
\]
independent uniform matrices. Conditional on
$\mathcal E$, they are therefore independent
uniform invertible matrices, and so are their
inverses $H,H'$. Each of the last two calls thus
fails with conditional probability at most
$10^{-6}$.

A union bound over the six calls completes the proof.
\end{proof}

The inputs in the lemma are defined by the ideal execution, in which preceding inversion calls are answered correctly. 

\paragraph{Passing from average to worst-case multiplication.}
Fix $\delta=10^{-6}$ and repeat the procedure independently $L=1800$ times. Verify every proposed product $Q$ using the standard randomized test for $Q=BC$~\cite{Freivalds1979}, and return the first candidate that passes. Use enough independent verification vectors to make the probability of a false acceptance over all trials at most $10^{-3}$. Each verification costs $\widetilde O(n^2)$ operations.

For every fixed pair $B,C$, the probability that none of the $L$ trials satisfies $\mathcal E$ is at most $\exp(-L/512)$. When $B,C$ are uniform and independent, Lemma~\ref{lem:ideal-inputs} bounds the probability that some trial satisfies $\mathcal E$ but has an incorrect ideal oracle answer by $6L\delta$. Consequently the average probability of an incorrect output is at most
\[
\exp(-1800/512)+6\cdot1800\cdot10^{-6}+10^{-3}<0.05.
\]
We have obtained an algorithm $\Mult$ satisfying $\Pr_{B,C}[\Mult(B,C)=BC]>0.95$, with running time $O(T(n)+T(2n)+n^2)$ up to an absolute multiplicative constant.

For arbitrary fixed $X,Y$, sample independent uniform $U,V\in\F_q^{n\times n}$ and return
\[
\Mult(X+U,Y+V)-\Mult(X+U,V)-\Mult(U,Y+V)+\Mult(U,V).
\]
Each queried pair is marginally a pair of independent uniform matrices, as in the usual self-reduction for matrix multiplication~\cite{BlumLubyRubinfeld1993}. By the union bound, with probability at least $1-4(0.05)=0.8$ all four products are computed correctly, in which case the expression equals $XY$. The algorithm can verify its proposed answer using Freivalds' test. Repeating with fresh randomness and retaining the first verified answer gives high success probability on every input, with a logarithmic overhead. This proves Proposition~\ref{prop:high-success}.

\paragraph{Simultaneous invertibility over every finite field.}
\begin{lemma}\label{lem:complementary}
For every finite field $\F_q$ and every $n\ge3$,
\[
\Pr_{R\sim\F_q^{n\times n}}[R,I-R\in\GL_n(\F_q)]\ge\frac1{32}.
\]
\end{lemma}
\begin{proof}
Write $r_i$ for the $i$th column of $R$ and expose the columns sequentially. Require both $r_1,\ldots,r_i$ and $e_1-r_1,\ldots,e_i-r_i$ to be linearly independent. Conditional on success through the first $i-1$ columns, the forbidden values for $r_i$ lie in a linear and an affine subspace, each of dimension $i-1$. The conditional success probability is therefore at least $1-2q^{i-1-n}$.

For $q\ge3$ we obtain
\[
\beta_n\ge\prod_{j=1}^n(1-2q^{-j})\ge(1-2/3)\left(1-\sum_{j=2}^\infty2\cdot3^{-j}\right)=\frac29.
\]
For $q=2$, expose first the initial $n-1$ columns and let $F$ be the event that both prefixes are linearly independent. The same argument gives $\Pr[F]\ge\gamma_{n-1}$. On $F$, the spans $U,V$ of the two prefixes are hyperplanes. If $U=V$, then $e_i=r_i+(e_i-r_i)\in U$ for all $i<n$, so $U=V=\operatorname{span}(e_1,\ldots,e_{n-1})$. This implies that the last coordinate of each of the first $n-1$ columns of $R$ is zero, an event of unconditional probability $2^{-(n-1)}$.

When $U\ne V$, the two forbidden sets for the last column are $U$ and $e_n-V$. These are distinct affine hyperplanes intersecting in a codimension-two affine subspace, so a quarter of all possible final columns avoid both. Hence
\[
\beta_n\ge\frac14\bigl(\gamma_{n-1}-2^{-(n-1)}\bigr).
\]
For $n\ge3$, the expression in parentheses is at least $\gamma_2-2^{-2}=1/8$, because $\gamma_k-2^{-k}$ is increasing for $k\ge2$. This completes the proof.
\end{proof}

\subsection{Amplification from any positive success probability}\label{sec:inversion-amplification}
We now show that an inverter with any positive average success probability can be transformed into one satisfying the hypothesis of Proposition~\ref{prop:high-success}. 
We do so via a sequence of random rank-one perturbations, each such step simultaneously allows us to recover the inverse of the original matrix, yet perturbs it a bit which is eventually sufficient to get into the set of matrices on which the algorithm of the previous subsection succeeds.

The spectrum of the rank-one walk below follows from the classical
spectrum of the bilinear forms graph~\cite{Delsarte78}; see
also~\cite[Section~7]{BCIM18}. We include a short Fourier-analytic
proof for completeness.

\begin{lemma}\label{lem:walk}
For $f:\F_q^{n\times n}\to[0,1]$, define
\[
(Pf)(X)=\E_{u,v\sim\F_q^n}[f(X+uv^\top)].
\]
If $\E_X [f(X)]=\alpha$, where $X$ is uniform, then for every integer $r\ge0$,
\[
\Var(P^r f)\le q^{-2r}\alpha.
\]
\end{lemma}
\begin{proof}
Fix a nontrivial additive character $\psi$ of $\F_q$, that is an additive homomorphism~$\F_q\rightarrow \mathbb C$ that is not constant (e.g., if~$q$ is prime then we can fix~$\psi(x)=e^{2\pi i x / q}$). 
The Fourier characters on the additive group of $n\times n$ matrices are $\chi_H(X)=\psi(\sum_{i,j}H_{ij}X_{ij})$. The eigenvalue of $P$ at the character $\chi_H$ is
\[
\E_{u,v}[\psi(u^\top Hv)]=\Pr_u[H^\top u=0]=q^{-\rank H}.
\]
All nonconstant eigenvalues are therefore at most $q^{-1}$. By Parseval's identity,
\[
\Var(P^r f)=\sum_{H\ne0}q^{-2r\rank H}|\widehat f(H)|^2\le q^{-2r}\Var(f)\le q^{-2r}\alpha,
\]
where the last inequality uses $0\le f\le1$.
\end{proof}

Intuitively, the above lemma implies that repeatedly adding a random rank-one matrix generically gets us ``closer'' to a uniform input from the perspective of any numerical test (including, whether the resulting matrix has a sufficiently high success probability in the average-case algorithm).

\begin{lemma}\label{lem:boost}
Suppose $\Inv_n$ correctly computes the inverse of a uniform matrix in $\GL_n(\F_q)$ with probability at least $\eps$. For any $0<\delta<1/2$, there is an algorithm $\Inv_n^+$ whose average success on $\GL_n(\F_q)$ is at least $1-\delta$ and whose running time is $\widetilde O_{\eps,\delta}(T(n)+n^2)$. For variable $\eps,\delta$, the overhead is near-linear in $1/\eps$ and logarithmic in $1/\delta$.
\end{lemma}
\begin{proof}
Define $f(X)$ to be the probability, over the randomness of $\Inv_n$, that $\Inv_n(X)=X^{-1}$ when $X$ is invertible, and set $f(X)=0$ otherwise. Its uniform mean $\alpha$ satisfies $\alpha\ge\eps\gamma_n\ge\eps/4$. By Lemma~\ref{lem:walk} and Chebyshev's inequality,
\[
\Pr_{X\sim\GL_n}\bigl[(P^r f)(X)<\alpha/2\bigr]\le\frac{4q^{-2r}}{\alpha\gamma_n}\le\frac{64q^{-2r}}{\eps}.
\]
Choose $r=\left\lceil\frac12\log_q\frac{192}{\eps\delta}\right\rceil$. Then, except for a $\delta/3$ fraction of invertible inputs, the probability that $\Inv_n$ correctly inverts $X+UV^\top$, for independent uniform $U,V\in\F_q^{n\times r}$, is at least $\alpha/2\ge\eps/8$.

Recall the Woodbury identity~\cite{Woodbury1950}: for an invertible matrix $Y$, provided that $I_r-V^\top Y^{-1}U$ is invertible,
\[
(Y-UV^\top)^{-1}
=Y^{-1}+Y^{-1}U(I_r-V^\top Y^{-1}U)^{-1}V^\top Y^{-1}.
\]

To invert a given matrix $X$, we repeatedly sample independent uniform matrices $U,V\in\F_q^{n\times r}$ and query $\Inv_n$ on $Y=X+UV^\top$. If we assume that the algorithm returns a matrix $J$ for which $I_r-V^\top JU$ is invertible, we can compute the candidate
\[
Z=J+JU(I_r-V^\top JU)^{-1}V^\top J.
\]
Whenever $J=Y^{-1}$ and $X$ is invertible, the matrix determinant lemma guarantees that $I_r-V^\top JU$ is invertible, and the Woodbury identity gives $Z=X^{-1}$. Since $U$ and $V$ have only $r$ columns, this computation takes $O(n^2r+nr^2+r^3)$ additional operations, without multiplying two general $n\times n$ matrices.

Verify a candidate $Z$ by checking $XZw=w$ for independent uniform vectors $w$. A wrong candidate passes a single check with probability at most $1/q$. Take $s=\lceil(8/\eps)\ln(3/\delta)\rceil$ trials, and in each trial use $d=\lceil\log_q(3s/\delta)\rceil$ verification vectors. For every nonexceptional invertible input, the probability that no trial produces a correct inverse is at most $\delta/3$, and the probability that any incorrect candidate is accepted is at most $\delta/3$. Including the exceptional input fraction gives average failure probability at most $\delta$. The cost is
\[
O\bigl(s(T(n)+n^2r+nr^2+r^3+dn^2)\bigr),
\]
which has the claimed bound.
\end{proof}

\begin{proof}[Proof of Theorem~\ref{thm:main-inverse}]
Apply Lemma~\ref{lem:boost} to the algorithms in dimensions $n$ and $2n$, with $\delta=10^{-6}$. They now satisfy the hypothesis of Proposition~\ref{prop:high-success}. Composing the reductions gives a worst-case matrix multiplication algorithm with running time $\widetilde O_{\eps}(T(n)+T(2n)+n^2)$. Its failure probability can be made inverse-polynomial by logarithmically many repetitions and verification.
\end{proof}

\section{Conditional reductions for rank and invertibility}\label{sec:rank}
This section is organized as follows:

First, we observe that by using known reductions we can show that every trapdoor matrix family also admits fast computation for both left and right multiplications, for computing its determinant, as well as for computing the inverse or solving a linear system. These additional quick operations are useful for our reductions in both this section and in Section~\ref{sec:det}.

Second, we give a simple unconditional reduction showing that deciding invertibility and exactly computing the rank are equivalent in the worst-case; hence, it is sufficient to construct a worst-case invertibility decision algorithm also for the case of rank computation.

Then, we design and analyze a simple type of query that will be used in both our reductions for rank and invertibility: it is a simple distributional test that we show, via a slightly technical linear-algebraic analysis, is sufficient to distinguish between an invertible and singular input matrix as long as the average-case algorithm gives any non-trivial answer. 

Finally, we plug this test into simple worst-case to average-case reductions for both invertibility and rank computation. 

\subsection{General capabilities of trapdoored matrices}\label{subsec:trapdoor}
Throughout the conditional results, we assume the existence of a trapdoored matrix family~\cite{VaikuntanathanZamir2026}: 
For a fixed finite field $\F_q$, a \emph{trapdoored-matrix family} is a family of samplers, indexed by the matrix dimension $n$ (which is also the security parameter), producing pairs $(R,\tau)$ with $R\in\F_q^{n\times n}$, such that:
(i) the marginal distribution of $R$ is computationally indistinguishable from the uniform distribution by every nonuniform probabilistic polynomial-time distinguisher, with negligible advantage in $n$;
(ii) sampling and writing the dense matrix take $\widetilde O(n^2)$ time; and
(iii) given $\tau$, the action $v\mapsto Rv$ can be evaluated exactly in $\widetilde O(n)$ time for every $v$.
The running times count field operations, and the pseudorandomness guarantee covers polynomially many independent samples by a hybrid argument. Vaikuntanathan and Zamir~\cite{VaikuntanathanZamir2026} construct such families over finite fields under corresponding field-specific LPN or McEliece-type assumptions. 

Applying $R$ to every column of a dense matrix computes $RA$ in $\widetilde O(n^2)$ time. Supporting right rather than left multiplication also follows: an independent sample $R$ provides a pseudorandom right mask $R^\top$, and each row of $AR^\top$ can be computed using the forward action of $R$. 

If the circuits provided by the trapdoor family are arithmetic (as is the case in all currently known construction), then we observe that supporting both right and left multiplication at the same time also holds generically:
When the trapdoor supplies a linear arithmetic circuit of size $u(n)$ computing $x\mapsto Rx$, the transposition principle also produces an $O(u(n)+n)$-size circuit for $R^\top y$ \emph{for that same matrix}. Indeed, we may differentiate $y^\top Rx$ with respect to $x$, and use the Baur--Strassen construction~\cite{BaurStrassen1983,BostanLecerfSchost2003,HhanLee2026} to quickly compute these partial derivatives as well. 

The forward action admits two further near-quadratic-time operations by classical black-box linear algebra: computing $\det R$, and solving $Rx=b$ for one prescribed vector $b$ when $R$ is invertible. Note we do not supply a near-linear-size circuit for the inverse map.

\begin{proposition}[Generic determinant and individual system solving from a fast forward action]\label{thm:trapdoor}
Let \(R\in\F_q^{n\times n}\) be given by a routine computing \(Rv\) in \(\widetilde O(n)\) time, together with an explicitly accessible dense representation or a sampler of total cost \(\widetilde O(n^2)\). Then \(\det R\) can be computed with high probability in \(\widetilde O(n^2)\) time. If \(R\) is invertible, \(R^{-1}b\) for one prescribed right-hand side can also be computed at that cost. 
\end{proposition}

The proposition follows from the next observation, which is a rephrasing of existing reductions by Wiedemann~\cite{Wiedemann1986}. Essentially, we use the trapdoor repeatedly to compute a sequence~$v,Av,A^2v,...,A^{2n}v$, then use quick algorithms to find a zeroing linear recurrence on it which is sufficient to recover~$A$'s minimal polynomial and in particular to recover both~$\det(A)$ and~$A^{-1}$.

\begin{lemma}[Black-box determinant and linear system solving]\label{lem:black-box}
Let $R\in\F_q^{n\times n}$ be represented by a routine that computes $Rv$ in time $\mu(n)$. There are randomized algorithms computing $\det R$ and, when $R$ is invertible, $R^{-1}b$ for one given $b\in\F_q^n$, using $\widetilde O(n)$ applications of the routine and $\widetilde O(n^2)$ additional field operations. Their failure probability can be made inverse-polynomial. 
\end{lemma}
\begin{proof}
We recall the black-box method of Wiedemann~\cite{Wiedemann1986}: For vectors $u,v\in\F_q^n$, consider the scalar sequence
\[
a_i=u^\top R^iv,\qquad i=0,1,\ldots,2n-1.
\]
The sequence satisfies a linear recurrence of order at most $n$, since the minimal polynomial of $R$ zeroes it. Its first $2n$ terms can be computed by starting from $v$ and repeatedly applying the routine for $R$, at a cost of $O(n)$ matrix-vector products and $O(n^2)$ additional operations. Then, the Berlekamp--Massey algorithm~\cite{Berlekamp1968,Massey1969} can be used to quickly recover the monic minimal recurrence polynomial in $O(n^2)$ field operations.

To obtain the determinant, work if necessary over an extension field $E/\F_q$ with $|E|\ge 20n^2$. Choose a random invertible diagonal multiplier $D$ over $E$. For a nonsingular $R$, Theorem~4.2 of Chen \emph{et al.}~\cite{ChenEberlyKaltofenSaundersTurnerVillard2002}, building on Wiedemann and Kaltofen--Saunders~\cite{Wiedemann1986,KaltofenSaunders1991}, states that $DR^\top$ is cyclic with probability at least $1-n(n-1)/|S|$ when the diagonal entries are independent uniform elements of a finite set $S$. Apply this with $S=E^\times$ and transpose: $RD$ is cyclic, meaning that its minimal and characteristic polynomials agree, with probability at least $1-n(n-1)/(|E|-1)$. Independent random projection vectors recover the full minimal polynomial from the scalar recurrence with another constant probability; for a cyclic matrix this follows by the usual polynomial-identity bound on cyclic input and output vectors. We repeat until Berlekamp--Massey returns a recurrence polynomial $p(t)$ of degree $n$, or until a logarithmic number of trials have failed.

Whenever $\deg p=n$, it must be the characteristic polynomial of $RD$, and hence
\[
 \det R=\frac{(-1)^np(0)}{\det D}.
\]
In particular, every \emph{nonzero} determinant returned in this way is correct, so the invertible-mask rejection step below has one-sided error: it never accepts a singular matrix. If no degree-$n$ recurrence is found, we return zero. This is correct on singular inputs, and fails with only inverse-polynomial probability on nonsingular inputs. Extending the action of $R$ from $\F_q$ to $E$ requires applying the original routine separately to the coordinates in an $\F_q$-basis of $E$. Since $[E:\F_q]=O(\log n)$, this costs only polylogarithmic overhead in field operations.

For a prescribed right-hand side $b$, use the same scalar-sequence method to obtain a zeroing polynomial $p(t)=p_0+p_1t+\cdots+p_dt^d$ for $R$, which is the minimal polynomial with high probability. Since $R$ is nonsingular, $p_0\ne0$. The identity $p(R)b=0$ gives
\[
R^{-1}b=-p_0^{-1}(p_1b+p_2Rb+\cdots+p_dR^{d-1}b).
\]
The right-hand side uses at most $n$ applications of $R$ to vectors and can be checked by computing $Rx$. Repeating independent trials gives high probability of success. Both algorithms therefore use $\widetilde O(n)$ matrix-vector products and $\widetilde O(n^2)$ additional field operations.
\end{proof}

\begin{proof}[Proof of Proposition~\ref{thm:trapdoor}]
Apply Lemma~\ref{lem:black-box} with $\mu(n)=\widetilde O(n)$. Sampling the matrix and trapdoor, computing its determinant, and solving any one prescribed invertible system all take $\widetilde O(n^2)$ time.
\end{proof}

\subsection{Equivalence of invertibility and rank computation}
We next make the simple unconditional observation that rank and invertibility are computationally equivalent. One direction is trivial (as rank implies whether a matrix is invertible), so we prove the converse direction.

\begin{proposition}[Random bordering]\label{prop:border}
For arbitrary $A\in\F_q^{n\times n}$ of rank $n-d$, independently sample uniform $U\in\F_q^{n\times k}$, $V\in\F_q^{k\times n}$ and $W\in\F_q^{k\times k}$, and form
\[
 B_k=\begin{pmatrix}A&U\\V&W\end{pmatrix}.
\]
If $k<d$, then $B_k$ is singular. If $k\ge d$, then
\[
 \Pr[B_k\in\GL_{n+k}]=\frac{(1-s_{q,k})^2}{(1-s_{q,k-d})}\ge\frac1{16}.
\]
Consequently, worst-case rank reduces unconditionally to worst-case nonsingularity with $O(\log^2 n)$ oracle calls on matrices of dimension at most $2n$, and $\widetilde O(n^2)$ additional work.
\end{proposition}
\begin{proof}
If $k<d$, adding $k$ columns and $k$ rows cannot bring rank $n-d$ to $n+k$. Otherwise change bases, only in the analysis, so that $A=\operatorname{diag}(I_{n-d},0_d)$. Eliminating the identity block leaves
\[
 \begin{pmatrix}0_d&U_0\\V_0&W'\end{pmatrix},
\]
where $U_0$ and $V_0$ are independent uniform $d\times k$ and $k\times d$ matrices, and $W'$ is independent uniform conditional on the other blocks. Invertibility requires full row rank of $U_0$ and full column rank of $V_0$, each with probability $(1-s_{q,k})/(1-s_{q,k-d})$. Conditional on these events, elimination leaves an independent uniform $(k-d)\times(k-d)$ block, invertible with probability $(1-s_{q,k-d})$. This proves the formula. Since $(1-s_{q,k})\ge1/4$, its value is at least $1/16$.
Repeated trials can then be used to test the monotone predicate $d\le k$ with arbitrarily small error. Finally, we use Binary search to determine $d$. 
Each query may be padded with an identity block to dimension exactly $2n$, so the bound uses only the oracle running time at $2n$. 
\end{proof}
Thus rank and invertibility are thus equivalent in a fine-grained manner.

\subsection{Rank profiles and masks of prescribed rank}

For $0\le j\le n$, let $\mathcal R_j$ denote the uniform distribution on matrices of rank $n-j$. Consider an arbitrary function $f:\F_q^{n\times n}\to[0,1]^m$, and define
\[
b_j=\E_{X\leftarrow\mathcal R_j}[f(X)].
\]
We call $(b_0,\ldots,b_n)$ the rank profile of $f$. Later, we will use $f(X)$ to represent the probability of an average-case algorithm on input $X$, or a vector containing the probabilities of all its possible outputs. For now, no assumption on $f$ is needed beyond its range.

We compare the values of $f$ on random matrices of prescribed rank with those obtained by randomly masking an arbitrary matrix. Set $D_j=\operatorname{diag}(I_{n-j},0_j)$, and let $P,Q,S$ be independent uniformly random invertible matrices. Since $PD_jS$ is uniform in $\mathcal R_j$, we have
\[
\E[f(PD_jS)]=b_j.
\]
For an arbitrary matrix $A$, consider instead the query
\[
PD_jQAS.
\]
If $A$ is invertible, this matrix is again uniform in $\mathcal R_j$, and its expected value under $f$ is $b_j$.

Now suppose that $\rank A=n-d$. Let $U$ be the image of $A$, which has dimension $n-d$, and let $W$ be the kernel of $D_jQ$, which is a uniformly random $j$-dimensional subspace. Then
\[
\rank(PD_jQAS)=n-d-\dim(U\cap W).
\]
Writing $K_{j,d}=d-j+\dim(U\cap W)$, the resulting rank is $n-j-K_{j,d}$. Equivalently, $K_{j,d}$ is the intersection dimension of a fixed $d$-dimensional subspace and a uniformly random $(n-j)$-dimensional subspace. Indeed, taking orthogonal complements gives
\[
\dim(U^\perp\cap W^\perp)=d-j+\dim(U\cap W),
\]
and $W^\perp$ is uniformly random among $(n-j)$-dimensional subspaces. Since the outer masks $P,S$ make the query uniform within its resulting rank class, its expected value is
\begin{equation}\label{eq:profile}
g_d(j)=\E[b_{j+K_{j,d}}].
\end{equation}
In particular, $g_0(j)=b_j$ for every $j$, while $g_d(0)=b_d$.

Thus, for an invertible matrix, the two queries have the same expected value for every $j$. For a singular matrix, their expectations are $b_j$ and $g_d(j)$, which may differ. 
We will use this difference to detect whether the matrix is invertible or not.
Testing only with $j=0$ is insufficient: we could have $b_d=b_0$ even when the profile is far from constant. The following lemma shows that whenever the profile differs sufficiently from $b_0$ on average, testing only $O(\log(1/\eta))$ values of $j$ suffices to distinguish every singular matrix.

For this statement, write
\[
\pi_j=\Pr_{X\sim\F_q^{n\times n}}[\rank X=n-j],
\]
and use the maximum norm $\|v\|_\infty=\max_i|v_i|$ for vector-valued profiles.

\begin{lemma}[Weighted rank-profile separation]\label{lem:separation}
Let $0<\eta\le1$, let $b_0,\ldots,b_n\in[0,1]^m$, and put
\[
J=\left\lceil\log_q\frac{100}{\eta}\right\rceil.
\]
Assume $n>J$ and
\[
\sum_{j=1}^n\pi_j\|b_j-b_0\|_\infty\ge\eta.
\]
Then, for every $1\le d\le n$, some $0\le j\le J$ satisfies
\[
\|g_d(j)-b_j\|_\infty>\eta/100.
\]
The constants are independent of the field size and the vector dimension $m$.
\end{lemma}

\begin{proof}
Fix $d$, and suppose to the contrary that
\[
\Delta:=\max_{0\le j\le J}\|g_d(j)-b_j\|_\infty\le\eta/100.
\]
We show that this forces the profile to be close to $b_0$ on average, contradicting the hypothesis. Since $g_d(0)=b_d$, we already have $\|b_d-b_0\|_\infty\le\Delta$. We first bound the differences between consecutive profile values at indices $j\ge d$, and then use these bounds to handle the remaining indices.

Write $e_t=b_{t+1}-b_t$ and define
\[
M=\max_{d\le t<n}q^{-t}\|e_t\|_\infty,
\]
with $M=0$ when $d=n$. The weighting allows us to control the contribution of later differences even when the intersection $K_{j,d}$ has dimension greater than one.

For $d\le j<n$, let $p_t=\Pr[K_{j,d}=t]$ and $p_+=\Pr[K_{j,d}\ge1]$. Counting subspaces with a prescribed intersection dimension gives, for consecutive feasible $t\ge1$,
\[
\frac{p_{t+1}}{p_t}
=q^{d-j-2t-1}
\frac{(1-q^{-(d-t)})(1-q^{-(n-j-t)})}
{(1-q^{-(t+1)})(1-q^{-(j+t-d+1)})}
\le\frac{q}{(q^2-1)^2}.
\]
Here $j\ge d$ and $t\ge1$, so the exponent in the first factor is at most $-3$, while both denominator factors are at least $1-q^{-2}$. The ratio bound implies that the probabilities of larger intersections decrease geometrically. More precisely, for every $a\ge1$,
\[
\Pr[K_{j,d}\ge a+1\mid K_{j,d}\ge1]
\le\left(\frac{q}{(q^2-1)^2}\right)^a.
\]
Consequently, since $q^2/(q^2-1)^2\le4/9$,
\[
\sum_{a\ge1}q^a\Pr[K_{j,d}\ge a+1\mid K_{j,d}\ge1]
\le\sum_{a\ge1}(4/9)^a=\frac45.
\]
We also need a lower bound on the probability of a nontrivial intersection. Any fixed nonzero vector in the $d$-dimensional subspace belongs to the random $(n-j)$-dimensional subspace with probability $(q^{n-j}-1)/(q^n-1)$. Therefore,
\[
p_+\ge\frac{q^{n-j}-1}{q^n-1}\ge\tfrac12q^{-j}.
\]

We now express the difference $g_d(j)-b_j$ using the successive differences $e_t$. The term $e_j$ appears whenever $K_{j,d}\ge1$, the term $e_{j+1}$ appears whenever $K_{j,d}\ge2$, and so on. Thus
\[
g_d(j)-b_j
=p_+e_j+\sum_{a\ge1}\Pr[K_{j,d}\ge a+1]e_{j+a}.
\]
For $d\le j\le J$, the assumption $\|g_d(j)-b_j\|_\infty\le\Delta$ and the preceding probability bounds give
\[
q^{-j}\|e_j\|_\infty\le2\Delta+\tfrac45M.
\]
For $j>J$, we instead use $\|e_j\|_\infty\le1$, which gives $q^{-j}\|e_j\|_\infty\le q^{-(J+1)}$. If the maximum defining $M$ is attained at an index at most $J$, the first inequality implies $M\le10\Delta$; otherwise the second bound applies. Hence
\begin{equation}\label{eq:weighted-maximum}
M\le\max\{10\Delta,q^{-(J+1)}\}.
\end{equation}

We next use this bound to compare $b_j$ with $b_0$. Recall that $\|b_d-b_0\|_\infty\le\Delta$. For $j\ge d$, summing the successive differences gives
\[
\|b_j-b_d\|_\infty
\le M\sum_{t=d}^{j-1}q^t
\le\frac{Mq^j}{q-1}.
\]

For $j<d$, we use the alternative description of the intersection variable from above. Let $U$ be a fixed $(n-d)$-dimensional subspace and $W$ a uniformly random $j$-dimensional subspace, and put $L=\dim(U\cap W)$. Then $K_{j,d}=d-j+L$, so
\[
g_d(j)=\E[b_{d+L}].
\]
We bound the possible increase from $b_d$ to $b_{d+L}$ by counting one-dimensional subspaces in $U\cap W$. An $L$-dimensional space contains $(q^L-1)/(q-1)$ such subspaces. Each one-dimensional subspace of $U$ belongs to $W$ with probability $(q^j-1)/(q^n-1)$. By linearity of expectation,
\[
\E\left[\frac{q^L-1}{q-1}\right]
=\frac{(q^{n-d}-1)(q^j-1)}{(q-1)(q^n-1)}
\le\frac{q^{j-d}}{q-1}.
\]
On the other hand, summing the differences between $b_d$ and $b_{d+L}$ gives
\[
\|b_{d+L}-b_d\|_\infty
\le M\sum_{t=d}^{d+L-1}q^t
=Mq^d\frac{q^L-1}{q-1}.
\]
Taking expectations, we obtain
\[
\|g_d(j)-b_d\|_\infty\le\frac{Mq^j}{q-1}.
\]
Together with $\|g_d(j)-b_j\|_\infty\le\Delta$ and $\|b_d-b_0\|_\infty\le\Delta$, this bounds the remaining indices $j<d$. Combining both cases, for every $0\le j\le J$ we have
\begin{equation}\label{eq:weighted-early}
\|b_j-b_0\|_\infty\le2\Delta+\frac{Mq^j}{q-1}.
\end{equation}

We have shown that if all the comparisons are close, then $b_j$ is close to $b_0$, with an error growing at most proportionally to $q^j$. We now average over the rank distribution of a uniform matrix. The factor $q^j$ causes no difficulty because its expectation under this distribution is less than two. Indeed, every fixed nonzero vector belongs to the kernel of a uniform matrix with probability $q^{-n}$, and therefore
\[
\sum_{j=0}^n\pi_jq^j
=\E_{X\sim\F_q^{n\times n}}[|\ker X|]
=1+(q^n-1)q^{-n}=2-q^{-n}<2.
\]
This also implies $\sum_{j>J}\pi_j\le2q^{-(J+1)}$. Averaging \eqref{eq:weighted-early} over $j\le J$, and using the trivial bound $\|b_j-b_0\|_\infty\le1$ for $j>J$, gives
\[
\sum_{j=1}^n\pi_j\|b_j-b_0\|_\infty
\le2\Delta+\frac{2M}{q-1}+2q^{-(J+1)}.
\]
By the choice of $J$, we have $q^{-(J+1)}\le\eta/(100q)$. Also, \eqref{eq:weighted-maximum} and $\Delta\le\eta/100$ imply $M\le\eta/10$. Since $q\ge2$, the last display is at most
\[
\frac{\eta}{50}+\frac{\eta}{5}+\frac{\eta}{100}<\eta,
\]
contradicting the hypothesis.
\end{proof}

\paragraph{Using the lemma algorithmically.}
Let~$A$ be a target input matrix with an unknown rank.
For each $0\le j\le J$, we estimate the expected responses to the calibration query $PD_jS$ and the target query $PD_jQAS$. Each coordinate is estimated to accuracy $\eta/1000$, and we declare the target singular if any empirical maximum-norm difference exceeds $\eta/200$. For an invertible target, the true expectations agree for every $j$; for any singular target, the lemma guarantees that at least one comparison differs by more than $\eta/100$. Thus the two cases are distinguished with high probability.

When the coordinates correspond to probabilities of different algorithm outputs, a single collection of samples estimates all coordinates simultaneously by recording their frequencies. With failure probability at most $\zeta$, $O(\eta^{-2}\log((m+1)(J+1)/\zeta))$ samples per query suffice. Since $J=O(\log(1/\eta))$, the total number of oracle calls is $\widetilde O(\eta^{-2})$ when $m$ is polynomial in $n$.

The analysis above uses genuinely uniform masks. Under the trapdoored-matrix assumption, independent pseudorandom masks implement the same queries in $\widetilde O(T(n)+n^2)$ time per sample, with negligible error from pseudorandom substitution. The reduction does not need to know the rank of the target matrix. If $n\le J$, direct elimination takes $O(n^3)\le O(Jn^2)$ operations, within the claimed overhead.

\subsection{Optimal success thresholds for invertibility and rank}
Recall that $\pi_j$ is the probability of rank $n-j$ for a uniform matrix. These probabilities are
\[
 \pi_j=q^{-j^2}
 \frac{(1-s_{q,n})^{\,2}}{(1-s_{q,j})^{\,2}(1-s_{q,n-j})},
 \qquad s_{q,0}=0.
\]
For $q\ge3$, invertibility is the most likely rank, so the best constant answers for Boolean invertibility and exact rank both have accuracy $1-s_{q,n}$. For $q=2$, the best Boolean answer is ``singular'', with accuracy $s_{2,n}$, whereas for $n\ge2$ the most likely rank is $n-1$, with probability $\pi_1=2(1-2^{-n})(1-s_{2,n})$.

\paragraph{Testing invertibility.}
Let $b_j$ be the probability that the Boolean oracle outputs ``invertible'' on a uniform rank-$(n-j)$ matrix, and abbreviate $s=s_{q,n}$. Its average accuracy is
\[
 \alpha=s+(1-s)b_0-\sum_{j\ge1}\pi_jb_j.
\]
A constant randomized predictor reporting ``invertible'' with probability $b_0$ has accuracy $s+(1-2s)b_0\le\max\{s,1-s\}$. Consequently, if the oracle exceeds this baseline by $\eps$,
\[
 \sum_{j\ge1}\pi_j(b_0-b_j)\ge\eps,
 \qquad\text{so}\qquad
 \sum_{j\ge1}\pi_j|b_j-b_0|\ge\eps.
\]
The weighted separation lemma directly gives the required test for every singular target, with $J=\lceil\log_q(100/\eps)\rceil$ and separation $\eps/100$.

\paragraph{Exact rank.}
For each $j$, let $b_j$ now be the vector of probabilities of all possible rank outputs $0,\ldots,n$. Let $b_j(a)$ denote its coordinate for output $a$. The rank oracle has accuracy
\[
 \alpha=\sum_{j=0}^n\pi_jb_j(n-j).
\]
A constant randomized predictor with output distribution $b_0$ succeeds with probability at most $\max_j\pi_j$. Hence an advantage of $\eps$ implies
\[
 \sum_{j\ge1}\pi_j\bigl(b_j(n-j)-b_0(n-j)\bigr)\ge\eps,
\]
and therefore $\sum_{j\ge1}\pi_j\|b_j-b_0\|_\infty\ge\eps$. The vector-valued separation lemma applies without choosing any output label in advance. The same samples estimate all coordinates of each $b_j$ and $g_A(j)$ by their empirical histograms.

\begin{proof}[Completion of Theorem~\ref{thm:rank-conditional}]
For both oracles, the weighted separation lemma and its statistical test decide worst-case invertibility using $\widetilde O(\eps^{-2})$ queries of dimension $n$, each implementable in $\widetilde O(T(n)+n^2)$ time under the trapdoor assumption. The calibration can be carried out independently of the input matrix~$A$. Replacing uniform masks by pseudorandom ones changes the success probability negligibly for polynomial-time oracles and reductions. For exact worst-case rank, apply Proposition~\ref{prop:border}, amplifying each invertibility decision as needed and using queries on matrices of dimension $2n$. The resulting total time is $\widetilde O_q(\eps^{-2}(T(2n)+n^2))$.
\end{proof}

\section{Conditional reductions for determinant}\label{sec:det}
We distinguish three regimes. First, knowing the determinant of a trapdoored mask makes the previously known high-success reductions work under the general trapdoor hypothesis. Second, accuracy just above the always-zero baseline is enough to recover worst-case \emph{rank}, but not automatically a nonzero determinant value. Third, any advantage over uniform guessing \emph{on invertible matrices} allows the determinant to be recovered modulo a small subgroup; the field's multiplicative-group structure then determines when exact recovery is possible.

\subsection{Exact determinant recovery using generic trapdoor matrices}\label{subsec:det-exact}

The generic known-determinant capability of Proposition~\ref{thm:trapdoor} recovers the improved exact-determinant thresholds from earlier work. The subgroup theorem in the following subsections generalizes them for all $q>2$ and strictly improves them for even $q\ge4$. 

The algorithm and proof in this section were alluded in~\cite{VaikuntanathanZamir2026} and sketched in~\cite{HhanLee2026}, we repeat them while filling the details for completeness. 

Let $R$ denote an independently sampled matrix from a pseudorandom trapdoored family. By Proposition~\ref{thm:trapdoor}, we may compute $\det R$ correctly with high probability in $\widetilde O(n^2)$ time. We can also form $RA$ in $\widetilde O(n^2)$ time by applying the fast matrix-vector multiplication routine to the columns of $A$. The one-sided determinant procedure allows rejection sampling of invertible masks without ever accepting a singular mask; its false-rejection probability is made negligible by repetition.

In the proofs below we first let $R$ be genuinely uniform, and then replace it by the trapdoor family. This replacement is valid by computational indistinguishability: for every fixed input matrix $A$, all events used in the analysis are testable by polynomial-time computations from the dense matrix $R$, including the oracle evaluation on $RA$ and the determinant of $R$. The fixed input $A$ can be hardwired into a nonuniform distinguisher. Consequently their probabilities change by a negligible amount under the pseudorandom substitution.

For $q>2$, exact recovery follows from the subgroup theorem in Section~\ref{subsec:det-subgroup} and its exact-recovery consequence in Section~\ref{subsec:field-consequences}. In particular, the classical $1/2+\eps$ invertible-input threshold is recovered (or subsumed) under the generic trapdoor family hypothesis without a separate proof. 
We thus only prove the binary threshold here, since over $\F_2$ the determinant is the invertibility bit and the subgroup formulation is vacuous.

\paragraph{The binary field.}
Now suppose $q=2$.  Sample a uniform matrix $R$, compute its determinant, and record whether
\[
\Det(RA)\ne\det R.
\]
If $A$ is invertible, $RA$ is uniform in $\F_2^{n\times n}$, so the probability of disagreement is at most $1-(s_{2,n}+\eps)=(1-s_{2,n})-\eps$.

If $A$ is singular, we claim that the disagreement probability is at least $(1-s_{2,n})$. To see this, condition on $RA$. After a change of basis, this fixes the restriction of $R$ to the image of $A$, which has dimension less than $n$, while the remaining columns of $R$ are uniform. If the fixed restriction is not injective, then $\det R=0$ with probability one. Otherwise, the remaining columns are independent uniform vectors; for the final column to complete an invertible matrix, it must escape the span of the preceding $n-1$ columns, an event of conditional probability $1/2$. Thus, in all cases,
\[
\Pr[\det R=1\mid RA]\le\frac12.
\]
For any value chosen by the oracle as a function of $RA$, the conditional probability of predicting $\det R$ incorrectly is at least $\Pr[\det R=1\mid RA]$. Averaging yields a disagreement probability of at least $\Pr[\det R=1]=(1-s_{2,n})$.

We may therefore distinguish singular from invertible inputs by estimating the disagreement probability to additive error less than $\eps/2$, using $O(\eps^{-2}\log n)$ independent trials. Return $1$ as the determinant when the empirical disagreement is below $(1-s_{2,n})-\eps/2$, and return $0$ otherwise. As before, pseudorandom trapdoor samples can replace uniform masks, with $\widetilde O(T(n)+n^2)$ time per trial. This proves the binary case of Corollary~\ref{thm:det-reduction}.

\subsection{Normalized queries}\label{subsec:normalized}
We now repeat the same simple algorithmic query distribution of the previous reductions, but phrase it generically enough to later use in a more elaborate manner.

Write $G=\F_q^\times$ and $m=q-1$ (including $m=1$ over $\F_2$). 
For a uniform $Y\in\GL_n(\F_q)$, define the \emph{normalized-answer distribution}
\begin{equation*}
 P=\operatorname{Law}\left(\frac{\Det(Y)}{\det Y}\right)
 \quad\text{on }\F_q.
\end{equation*}
In particular, $P(1)$ is the success probability of $\Det$ conditional on a uniform invertible input. Above the invertibility-only benchmark, $P(1)\ge1/m+\eps$, which forces a nontrivial multiplicative bias. For an arbitrary matrix $A$, sample $R$ uniformly from $\GL_n(\F_q)$ and define
\[
 Z_A=\frac{\Det(RA)}{\det R},\qquad P_A=\operatorname{Law}(Z_A).
\]
For $a\in G$ and a distribution $Q$ on $\F_q$, write $aQ$ for the distribution of $aW$ when $W\sim Q$; zero is fixed by this action.

\begin{lemma}[Shift versus uniformity]\label{lem:normalized}
If $A$ is invertible with determinant $d$, then $P_A=dP$. If $A$ is singular, then for every $z\in G$,
\[
 P_A(z)\le\frac1m.
\]
Indeed, in the singular case $P_A$ is a mixture of an atom at zero and the uniform distribution on $G$.
\end{lemma}
\begin{proof}
When $A$ is invertible, $Y=RA$ is uniform on $\GL_n$ and $\det R=\det Y/d$. Hence $Z_A=d\Det(Y)/\det Y$, proving $P_A=dP$.

Suppose that $A$ is singular. For each fixed $B=RA$, the determinant of $R$ is uniform on $G$ even after conditioning on $B$. To see this, for any $a\in G$ choose an invertible matrix $S$ which fixes the image of $A$ pointwise and has $\det S=a$. The bijection $R\mapsto RS$ leaves $RA$ unchanged and multiplies $\det R$ by $a$. Conditional on $RA$ and on the independent random coins of $\Det$, the numerator $\Det(RA)$ is fixed, while $\det R$ is uniform on $G$. If the numerator is nonzero, their ratio is uniform on $G$; otherwise it is zero. This proves the claim.
\end{proof}

\subsection{Rank recovery above the always-zero baseline}\label{subsec:det-to-rank}

Suppose that $\mathcal D$ computes the determinant of a uniform matrix correctly with probability at least $s_{q,n}+\eps$. For each $j$, let
\[
u_j=\Pr_{X\leftarrow\mathcal R_j}[\mathcal D(X)\ne0].
\]
Recall that $P(1)$ is the success probability of $\mathcal D$ on uniform invertible matrices, while $u_0$ is its probability of returning a nonzero value on such matrices. Thus $u_0/(q-1)$ is the average probability assigned by $P$ to a nonzero field element.

Since the determinant of a singular matrix is zero, the assumed success probability gives
\[
\begin{aligned}
\eps
&\le (1-s_{q,n})P(1)-\sum_{j\ge1}\pi_j u_j\\
&=(1-s_{q,n})\left(P(1)-\frac{u_0}{q-1}\right)
+\left(\frac{1-s_{q,n}}{q-1}-s_{q,n}\right)u_0
+\sum_{j\ge1}\pi_j(u_0-u_j)\\
&\le (1-s_{q,n})\left(P(1)-\frac{u_0}{q-1}\right)
+\sum_{j\ge1}\pi_j(u_0-u_j),
\end{aligned}
\]
where the last inequality follows from $s_{q,n}\ge1/q$.

Consequently, at least one of the following holds:
\[
P(1)-\frac{u_0}{q-1}\ge\frac{\eps}{2}
\qquad\text{or}\qquad
\sum_{j\ge1}\pi_j(u_0-u_j)\ge\frac{\eps}{2}.
\]
In the first case, $P$ is noticeably nonuniform on the nonzero field elements, unlike what would happen for a singular input (where we showed that this distribution is uniform), so this can be detected using a simple ``flatness'' test. 
In the second case, the function indicating whether $\mathcal D$ returns a nonzero value has a rank profile satisfying the hypothesis of Lemma~\ref{lem:separation}.
We next present a simple algorithm that distinguishes between invertible and singular matrices in each of the cases.

\paragraph{Case 1: A nonuniformity test for normalized answers.}
For any distribution $Q$ on $\F_q$ define
\[
 F(Q)=\max_{z\in\F_q^\times}
 \left|Q(z)-\frac{1-Q(0)}m\right|.
\]
For an input matrix $A$, use the normalized masking distribution $P_A$ of Lemma~\ref{lem:normalized}, obtained from $\Det(RA)/\det R$.
If $A$ is nonsingular with determinant $d$, then $P_A=dP$, so $F(P_A)=F(P)$. If $A$ is singular, Lemma~\ref{lem:normalized} gives $F(P_A)=0$. 

Calibrate $F(P)$ to additive error at most $\eps/32$. 
If the empirical value is at least $\eps/8$, estimate $F(P_A)$ to the same accuracy and declare $A$ invertible precisely when it exceeds $3\eps/64$. 
In this branch the true value $F(P)$ is at least $3\eps/32$, while every singular input has $F(P_A)=0$. Each normalized sample costs near-quadratic time using a trapdoored mask with known determinant.

\paragraph{Case 2: a nonconstant rank profile.}
Otherwise, the empirical flatness is below $\eps/8$, so $F(P)<5\eps/32<\eps/2$. Since $p-\mu\le F(P)$, the first alternative in our dichotomy cannot hold. Therefore,
\[
\sum_{j\ge1}\pi_j(u_0-u_j)\ge\eps/2.
\]
Thus the function indicating whether $\mathcal D$ returns a nonzero value has a rank profile satisfying the hypothesis of Lemma~\ref{lem:separation}, with $\eta=\eps/2$ and can be detected by the corresponding algorithm.
This also requires only $\widetilde O(\eps^{-2})$ oracle samples. 

These two calibration-selected branches complete the proof of Theorem~\ref{thm:det-hierarchy}(i). For $q=2$, the implication to worst-case rank follows already from Theorem~\ref{thm:rank-conditional}: the Boolean predictor $\boldsymbol1[\Det(X)\ne0]$ has accuracy at least the determinant accuracy, and the Boolean baseline is $s_{2,n}$. The dichotomy above is needed for $q\ge3$.

\subsection{Subgroup recovery above the invertibility baseline}\label{subsec:det-subgroup}

We have shown that average-case determinant success $s_{q,n}+\eps$ suffices for worst-case rank computation. We next show that a stronger success guarantee, exceeding random guessing among the nonzero field elements on uniformly random invertible matrices, allows us to recover the determinant itself up to a small multiplicative subgroup of $\F_q^\times$.

From now on let $q>2$, put $G=\F_q^\times$, and suppose the success probability $p=P(1)$ of $\Det$ on uniform invertible matrices satisfies
\[
 p\ge\frac1{q-1}+\eps.
\]
No correctness guarantee on singular inputs is required. If instead we are given the overall success bound $s_{q,n}+(1-s_{q,n})/(q-1)+\eps$, the same hypothesis follows, since conditional success is at least $1/(q-1)+\eps/(1-s_{q,n})$. This is exactly where the invertibility-only predictor stops being sufficient to explain the oracle's accuracy. The reduction first learns an ambiguity subgroup using \emph{only} samples independent of the target input matrix, and then uses that subgroup for all input matrices.

\paragraph{Learning a robust ambiguity subgroup.}\label{subsec:subgroup}
The exact multiplicative stabilizer
\[
 H_*(P)=\{a\in G:aP=P\}
\]
is a subgroup, which consists exactly of the multipliers that the entire distribution of the normalized test is invariant under.
Moreover, $|H_*(P)|P(1)\le1$ (as~$P(1)=P(a)$ for every~$a\in H_*(P)$), so $|H_*(P)|\le 1/P(1)<m$. Theoretically, the orbit $dP$ identifies $d$ up to $H_*(P)$. Algorithmically, however, $P$ may be exponentially close to an additional symmetry, making its exact stabilizer too expensive to learn. 
We thus use a robust, efficiently learnable supergroup of all near-symmetries needed for the recovery procedure.

Denote by
\[
 L=\left\lceil\frac2\eps\right\rceil,\qquad
 \rho=\frac{\eps^2}{10000}.
\]
Take independent samples of $P$ and construct an empirical distribution $\widehat P$ such that
\begin{equation}\label{eq:empirical}
 \|\widehat P-P\|_\infty\le\rho.
\end{equation}
For fixed $q$, $O(\rho^{-2}\log n)$ samples suffice for inverse-polynomial failure probability by Hoeffding's inequality; for non-constant $q$, an additional $O(\log q)$ factor in the sample count suffices. Here and below $\|Q-Q'\|_\infty=\max_{z\in\F_q}|Q(z)-Q'(z)|$.

Define
\begin{equation}\label{eq:near-symmetry}
 S=\left\{a\in G:\ \widehat P(a)\ge\frac\eps2
 \quad\text{and}\quad\|\widehat P-a\widehat P\|_\infty\le6\rho\right\},
 \qquad H=\langle S\rangle.
\end{equation}
One need not enumerate $G$ to find $S$: every candidate is a heavy atom of $\widehat P$, so there are at most $2/\eps$ of them. For each such candidate, the $\ell_\infty$ distance is evaluated by inspecting the union of the empirical supports of $\widehat P$ and $a\widehat P$. The group generated by $S$ can be enumerated by breadth-first search in its Cayley graph, as the next lemma proves it has constant size for fixed $\eps$.

\begin{lemma}[A small proper ambiguity subgroup]\label{lem:robust-subgroup}
On the event~\eqref{eq:empirical}, the subgroup $H$ of~\eqref{eq:near-symmetry} is proper and satisfies
\[
 \boxed{|H|<\frac{1}{1/m+\eps/2}}
 \quad\text{and hence}\quad
 |H|\mid m,\qquad |H|<\min\{m,2/\eps\}.
\]
More precisely, if $p=P(1)$, then $|H|<(p-\eps/4)^{-1}$.
\end{lemma}
\begin{proof}
For every $a\in S$, the empirical approximation gives
\[
 \|P-aP\|_\infty\le8\rho.
\]
The same bound holds for $a^{-1}$, and the triangle inequality is invariant under shifts. Consequently, if $g$ is a word of length $t$ in $S\cup S^{-1}$, then
\begin{equation}\label{eq:word-bound}
 \|P-gP\|_\infty\le8t\rho,
 \qquad P(g)\ge P(1)-8t\rho.
\end{equation}
For $t\le L$, our choices imply $8t\rho\le8L\rho<\eps/4$, and so
\[
 P(g)>p-\eps/4\ge\frac1m+\frac{3\eps}4.
\]
Suppose that $|H|>L$. Since $S\cup S^{-1}$ generates $H$, at least $L+1$ distinct group elements can be reached from the identity by words of length at most $L$. The displayed lower bound would give total mass greater than $(L+1)3\eps/4>1$, a contradiction. Thus $|H|\le L$.

Every $g\in H$ now has a word of length at most $|H|-1\le L$, so $P(g)>p-\eps/4$. Summing over $g\in H$ yields
\[
 1\ge\sum_{g\in H}P(g)>|H|(p-\eps/4)
 \ge |H|\left(\frac1m+\frac{3\eps}4\right).
\]
In particular $|H|<(p-\eps/4)^{-1}<(1/m+\eps/2)^{-1}$. The remaining statements follow from Lagrange's theorem and the strict inequalities $1/(1/m+\eps/2)<m,2/\eps$.
\end{proof}

\paragraph{Recovering the determinant coset.}\label{subsec:coset}
We next finish the proof of Theorem~\ref{thm:det-hierarchy}(ii). The calibration of $P$ and the choice of $H$ are performed independently of the input matrix $A$. For this input, sample and construct $\widehat P_A$ such that
\[
 \|\widehat P_A-P_A\|_\infty\le\rho
\]
with high probability. Search among the nonzero observed values $z$ for one satisfying
\begin{equation}\label{eq:coset-test}
 \widehat P_A(z)\ge\frac\eps2,
 \qquad \|\widehat P_A-z\widehat P\|_\infty\le2\rho.
\end{equation}
If no such $z$ exists, return zero and declare $A$ singular. Otherwise, output $zH$, or equivalently output $z^{|H|}$.

If $A$ is invertible, put $d=\det A$. Since $P_A=dP$ by Lemma~\ref{lem:normalized}, the candidate $z=d$ passes~\eqref{eq:coset-test}: its empirical frequency is at least $P(1)-\rho>\eps/2$, and the distance is at most $2\rho$. Thus the search succeeds.

Conversely, if $z$ passes for an invertible $A$, then
\[
 \|P-(z/d)P\|_\infty
 =\|P_A-zP\|_\infty\le4\rho.
\]
In particular, $P(z/d)\ge P(1)-4\rho$, hence $\widehat P(z/d)\ge P(1)-5\rho>\eps/2$. Also
\[
 \|\widehat P-(z/d)\widehat P\|_\infty\le6\rho.
\]
Therefore $z/d\in S\subseteq H$, proving that every accepted candidate yields the correct coset $zH=dH$. Since the multiplicative group of a finite field is cyclic, the kernel of $x\mapsto x^{|H|}$ is exactly $H$, so $z^{|H|}=d^{|H|}$.

If $A$ is singular, Lemma~\ref{lem:normalized} gives $P_A(z)\le1/m$ for every $z\in G$. But the distribution $zP$ has mass $P(1)\ge1/m+\eps$ at $z$, so $\|P_A-zP\|_\infty\ge\eps$. The empirical errors are too small for~\eqref{eq:coset-test} to hold. Hence the procedure declares singularity correctly.

The sample sizes are polynomial in $1/\eps$ and logarithmic in $n$ (and in $q$ if it varies); construction of the empirical distributions, verification of their shifts, and enumeration of $H$ require only comparable work. Every oracle sample can be generated using the trapdoored mask family and its known determinant in $\widetilde O(n^2)$ overhead. First analyze all polynomially many samples as if their masks were uniform in $\GL_n$; rejection sampling from the pseudorandom family (with its determinant computed as in Proposition~\ref{thm:trapdoor}) replaces each such mask with negligible change to any event decidable in polynomial time. A standard hybrid argument applies equally to the adaptive calibration and coset tests, since all outputs and the correct determinant of the fixed target are polynomial-time testable for the purpose of the indistinguishability proof. The subgroup calibration and coset recovery require polynomially many samples in $1/\eps$, as do the rank-profile tests. This establishes the claimed $\widetilde O_{q,\eps}(T(n)+n^2)$ running time and high success probability.

\subsection{Exact recovery and the role of the field}\label{subsec:field-consequences}

The preceding reduction recovers $\det(A)$ up to multiplication by a subgroup $H$ of $\F_q^\times$. The subgroup may depend on the average-case algorithm, but its size is bounded in terms of the algorithm's success probability. We now use this bound to determine when the reduction recovers the determinant exactly, or up to a prescribed subgroup.

Let $K<\F_q^\times$ be a subgroup, and consider the smallest possible order of an element outside $K$:
\[
r_K=\min\{\operatorname{ord}(a):a\in\F_q^\times\setminus K\}.
\]
If the subgroup $H$ recovered by our reduction has order smaller than $r_K$, then necessarily $H\subseteq K$. Thus recovering the determinant modulo $H$ also determines it modulo $K$.

\begin{corollary}[Recovery modulo a prescribed subgroup]\label{cor:target-subgroup}
Let $K<\F_q^\times$ be a subgroup. If the determinant oracle succeeds on a fraction at least $1/r_K+\eps$ of uniform invertible matrices, then, under the trapdoor assumption, the reduction recovers $\det(A)K$ on every input, with the same running-time exponent.
\end{corollary}
\begin{proof}
The proof of Lemma~\ref{lem:robust-subgroup} gives a subgroup $H$ with
\[
|H|<(p-\eps/4)^{-1}<r_K.
\]
If $H$ contained some $a\notin K$, then the order of $a$ would be at least $r_K$. By Lagrange's theorem, this order divides $|H|$, a contradiction. Hence $H\subseteq K$, and the computed coset $\det(A)H$ determines $\det(A)K$.
\end{proof}

\paragraph{Exact determinant recovery.}
Taking $K=\{1\}$ gives exact recovery. Since $\F_q^\times$ is cyclic of order $q-1$, the smallest order of a nonidentity element is the smallest prime divisor $\ell$ of $q-1$. Consequently, for every $q>2$, correctness on a fraction $1/\ell+\eps$ of uniform invertible matrices suffices for exact worst-case determinant computation. In terms of success on uniformly random matrices, it suffices to achieve
\[
\boxed{s_{q,n}+\frac{1-s_{q,n}}{\ell}+\eps.}
\]

For odd $q$, we have $\ell=2$, recovering the previously known $(1+s_{q,n})/2+\eps$ threshold (but with the weaker trapdoored family assumption). For fields of characteristic two with $q\ge4$, the number $q-1$ is odd, so $\ell\ge3$ and the threshold is strictly smaller. In particular, it can fall well below $1/2$, whereas all previous sufficient thresholds for exact determinant computation exceeded $1/2$.

When $q-1$ is itself prime, we have $\ell=q-1$, and the threshold coincides exactly with the anything-better-than-invertibility baseline. This holds, for example, over $\F_4,\F_8,\F_{32}$, and $\F_{128}$, and would hold for infinitely many fields if there are infinitely many Mersenne primes. However, the threshold can be made arbitrarily small over infinitely many finite fields without any number-theoretic conjecture.

\begin{proof}[Proof of Corollary~\ref{cor:small-success}]
Let $r$ be an odd prime and set $q=2^r$. We show that every prime divisor $\ell$ of $q-1$ is at least $2r+1$.

Since $2^r\equiv1\pmod\ell$, the multiplicative order of $2$ modulo $\ell$ divides $r$. As $r$ is prime and $2\not\equiv1\pmod\ell$, this order must be $r$. Therefore $r$ divides $\ell-1$. Both $r$ and $\ell$ are odd, so in fact $2r$ divides $\ell-1$, giving
\[
\ell\ge2r+1.
\]

On the other hand, the probability that a uniform matrix over $\F_q$ is singular satisfies, for every matrix dimension $n$,
\[
s_{q,n}<\sum_{i=1}^{\infty}q^{-i}=\frac1{q-1}.
\]
The exact-recovery threshold established above, excluding its arbitrarily small positive slack, is therefore at most
\[
s_{q,n}+\frac{1-s_{q,n}}{\ell}
<\frac1{2^r-1}+\frac1{2r+1}=O(1/r).
\]
This tends to zero as $r$ ranges over the infinitely many odd primes. Thus, for every fixed $\eta>0$, we can choose such a field with the displayed threshold below $\eta/2$, and apply the reduction with slack $\eta/2$. The field is fixed as the matrix dimension grows, and the reduction assumes the corresponding trapdoored-matrix family over that field.
\end{proof}

Some numerical examples are given in Table~\ref{tab:even-fields}. The field $\F_{2048}$, for instance, has $2047=23\cdot89$, and hence exact determinant computation follows already from average success slightly above $4.4\%$, compared with the previous threshold of slightly above $50\%$.

\begin{table}[h]
\centering\small
\begin{tabular}{@{}rrrr@{}}
\toprule
$q$ & Smallest prime $\ell\mid(q-1)$ & New threshold & Previous threshold \\
\midrule
4 & 3 & $0.54+\eps$ & $0.66+\eps$ \\
8 & 7 & $0.26+\eps$ & $0.57+\eps$ \\
16 & 3 & $0.38+\eps$ & $0.53+\eps$ \\
32 & 31 & $0.063+\eps$ & $0.52+\eps$ \\
64 & 3 & $0.34+\eps$ & $0.51+\eps$ \\
2048 & 23 & $0.044+\eps$ & $0.5002+\eps$ \\
\bottomrule
\end{tabular}
\caption{Illustrative limiting success thresholds for exact determinant computation as $n\to\infty$, over fixed fields of characteristic two. Entries are rounded. The reductions assume the corresponding trapdoored-matrix families.}\label{tab:even-fields}
\end{table}

\paragraph{Recovery up to sign.}
The same argument applies when exact recovery is relaxed to recovery up to a multiplicative subgroup. For example, over fields of odd characteristic, taking $K=\{1,-1\}$ gives recovery up to sign whenever the conditional success probability exceeds $1/r_K$. This can be substantially smaller than the threshold for exact recovery, depending on the multiplicative structure of the field.
For example, when $q=2p+1$ is a safe prime with $p$ odd, we have $r_K=p$, and recovery up to sign follows from success greater than $2/(q-1)$ on uniform invertible matrices. Thus, if infinitely many safe primes exist, as conjectured, then for every fixed $\eps>0$, there are infinitely many prime fields over which success $1/(q-1)+\eps$ suffices for recovery up to sign.

\section{Discussion and Open Problems}\label{sec:discussion}

While we settle the worst-case to average-case reduction picture for matrix inversion, we leave a couple of major open problems for the other operations.

\paragraph{Overcoming subgroup ambiguities for determinant.}
Our determinant reductions use multiplicative masking, which cannot distinguish determinant values that differ by an element of the ambiguity subgroup. This limitation is closely related to an open question about the worst-case complexity of determinant computation. For any subgroup $H\le\F_q^\times$ of order $h$, consider an algorithm that returns zero on singular matrices and a fixed representative of $\det(A)H$ otherwise. This algorithm succeeds on a fraction
\[
s_{q,n}+\frac{1-s_{q,n}}{h}
\]
of uniformly random matrices, but its output can be computed from $(\det A)^h$. Consequently, a worst-case to average-case reduction for exact determinant computation that worked at this success probability would also give an exponent-preserving worst-case reduction computing $\det A$ using an oracle for $(\det A)^h$.

Two notable cases are $h=2$, corresponding to recovering the determinant up to sign in odd characteristic, and $h=q-1$, corresponding to deciding invertibility. In particular, it remains open whether a fast worst-case algorithm for deciding invertibility can be used to compute the determinant with the same running-time exponent. This question is also connected to obtaining optimal success thresholds for our average-case reductions: we have shown that average-case determinant success $s_{q,n}+\eps$ already suffices for worst-case rank computation, and that worst-case rank and invertibility are equivalent. Therefore, an exponent-preserving worst-case reduction from determinant to invertibility would also establish a worst-case to average-case reduction for exact determinant computation at the optimal threshold $s_{q,n}+\eps$, under the same trapdoor assumption. More generally, understanding whether the multiplicative subgroup ambiguities can be removed would thus lead to several additional breakthroughs. 

\paragraph{Unconditional reductions for other linear-algebraic problems.}
Our main result establishes an unconditional worst-case to average-case reduction for matrix inversion. The reductions for rank, invertibility, and determinant, however, still rely on the existence of pseudorandom trapdoored matrices. These matrices allow us to apply random-looking multiplicative masks to arbitrary worst-case inputs in near-quadratic time. Using genuinely uniform dense matrices for the same purpose would generally require matrix multiplication, losing the desired running-time bound.

It remains open whether the conditional reductions for rank, invertibility, and determinant can be made unconditional while preserving their success thresholds and running-time exponents. Another important case is solving a single linear system $Ax=b$, for which the corresponding unconditional worst-case to average-case reduction is also unknown.

\begingroup
\bibliographystyle{alpha}
\bibliography{linear_algebra_reductions}
\endgroup
\end{document}